\documentclass[twocolumn]{aastex631} %linenumbers
\usepackage{needspace}

\newcommand{\be}{\begin{eqnarray}}
\newcommand{\ee}{\end{eqnarray}}
\newcommand{\bi}{\begin{itemize}}
\newcommand{\ei}{\end{itemize}}
\newcommand{\pomega}{{\varpi}}
\newcommand{\R}{{\cal R}}
\newcommand{\Q}{{Q_{\rm ice}}}

\begin{document}

\title{Enceladus's Limit Cycle}

\correspondingauthor{Yoram Lithwick}
\email{y-lithwick@northwestern.edu}

\author{Peter Goldreich}
\affiliation{California Institute of Technology, 1200 East California Boulevard, Pasadena, CA 91125}

\author[0000-0003-4450-0528]{Yoram Lithwick}
\affiliation{Department of Physics \& Astronomy, Northwestern University, Evanston, IL 60202, USA}
\affiliation{Center for Interdisciplinary Exploration \& Research in Astrophysics (CIERA), Evanston, IL 60202, USA}

\author{Jing Luan}

\begin{abstract}

Enceladus
exhibits some remarkable phenomena, 
 including
 water geysers spraying through surface cracks, 
  a global ice
 shell  that is librating atop an  ocean,
 a large luminosity, and
 rapid outward orbital migration. 
Here we model the coupled evolution of Enceladus's orbit and interior structure. We
 find that Enceladus is driven into a periodic state---a limit cycle.
  Enceladus's observed phenomena
 emerge  from the model, and
 the  predicted values for the orbital
eccentricity, libration amplitude, shell thickness, and luminosity  agree with observations.
A single limit  cycle lasts around ten million years, and  has three distinct stages: (1) freezing, (2) melting, and (3) resonant libration.
Enceladus is currently in the freezing stage, meaning that its ice shell is getting thicker.
That pressurizes the ocean, 
which in turn cracks the shell
 and pushes water up
through the cracks.
In this stage the orbital
eccentricity increases, 
     as Saturn pushes Enceladus
deeper into resonance with  Dione. 
Once the eccentricity is sufficiently high, tidal heating  begins to melt the shell, which is the 
second stage of the cycle.
In the third stage 
 the shell remains close to 3km thick.
 At that
  thickness the shell's natural libration
 frequency is  resonant with the orbital frequency.
The shell's librations are consequently driven to  large amplitude,
 for millions of years.
Most of the tidal heating of Enceladus occurs during this stage, and
%Some
%of the energy is stored up, to be   emitted later during the next freezing stage.
the
 observed luminosity is  a relic
from the last episode of resonant libration.

\end{abstract}

\section{Introduction}
\label{sec:intro}

 Enceladus is a small moon of Saturn, 
 with a radius of 252km. 
 It
  is covered by an ice shell,
  which conceals
 a global   ocean below 
    \citep{2016Icar..264...37T}. 
Near the south pole, water vapor and
frozen mist spray
from the ocean 
 through four long and deep parallel fractures in the shell
  \citep{2006Sci...311.1393P}, in 
  the form 
  of $\sim 100$ geysers \citep{2014AJ....148...45P}.
These observations, and many others, were gathered by the {\it Cassini} mission, from 2005-2017.
   The wealth of Enceladus's observed phenomema provides valuable clues into how tides
 operate, both in Enceladus and in Saturn. 
 See, e.g.,   \cite{2023SSRv..219...57N} and
 \cite{2024SSRv..220...20C}
 for  reviews.

  Enceladus's  ocean is prevented from  freezing over  by tidal heating within Enceladus, 
which is  driven by the   eccentricity  of
its orbit around Saturn. The current eccentricity is
\be
e_{\rm obs} &=& 0.0047  \ . \label{eq:eobs}
\ee
  Tidal heating should   quickly  circularize
the orbit.  But circularization is prevented by
 tides on Saturn,  which push Enceladus deeper into its 2:1 mean motion resonance (MMR) with its
 outer partner, Dione, 
 thereby
 maintaining a resonantly forced eccentricity for Enceladus.

The leading mechanism for tidal pushing by Saturn used to be  equilibrium 
 tides. 
 In the equilibrium tide scenario, Enceladus raises a tidal bulge in Saturn, and
 the time-varying bulge is presumed to dissipate energy, at a rate
 parameterized by an unknown tidal quality factor in Saturn ($Q_S$). 
 But recent observations found 
 that  Saturn's moons are migrating outward
 at rates that are  inconsistent  with this scenario \citep{2012ApJ...752...14L,2020NatAs...4.1053L}.
Lainey et al find that the migration timescale is 
\be
\tau_{\rm obs}\sim 10 {\rm Gyr} \ ,
\label{eq:tauobs}
\ee
for six of the moons, including Enceladus.
In contrast, the equilibrium tide scenario
 predicts a migration timescale  that increases  rapidly with distance from Saturn, 
 and so should be extremely long for the far-out moons, especially Rhea and Titan. 
 Rhea's well-observed migration rate is particularly constraining, and 
would imply $Q_S\approx 300$ under equilibrium tides. That  $Q_S$ is not only
 extremely small  for a fluid body such as Saturn, but  is also inconsistent with the values of $Q_S$ implied by other moons. 
\cite{2016MNRAS.458.3867F} showed that the fast observed migration rates could be understood 
if the moons were being pushed out by the ``resonance locking'' mechanism, rather than equilibrium tides.  In  resonance
 locking, a moon excites a near-resonant
  oscillation mode of Saturn. 
 Dissipation of the mode's energy
  forces the moon to maintain 
 an orbital frequency slightly less than the mode's frequency.
 As Saturn evolves,  the
 mode's frequency is presumed to decrease, on Gyr
 timescales. That forces the moon's frequency
 to decrease too, implying outward migration.
   We  adopt the resonance locking
 mechanism in this paper.

  Observations of the
     rotation state of Enceladus's surface find that it is librating on the timescale of its orbit. %\citep{2016Icar..264...37T}, 
     Its forced libration amplitude  
    is found to be
    \be
    \gamma_{\rm obs} \approx 0.1^\circ
    %\pm 0.009^\circ \ \ \ (3{\rm -}\sigma) 
    \label{eq:gammaobs}
    \ee
     with $\sim 20\%$  error bars \citep{2016Icar..264...37T,2024JGRE..12908054P}.
     This libration is forced by Saturn, as Saturn traces out its epicycle as
seen from Enceladus.
    But $\gamma_{\rm obs}$ is too large for
     an entirely solid moon. 
    Instead, there must be  a thin ice shell that is librating on top of 
    a global ocean.  From $\gamma_{\rm obs}$, the inferred average thickness of the shell is
    \be
     d_{\rm obs}\sim 20{\rm -}30 \ {\rm km} \ ,  \label{eq:dobs}
     \ee
  \citep{2016Icar..264...37T,2016GeoRL..43.5653C,2024JGRE..12908054P}. 
    Measurements of Enceladus's quadrupolar gravity \citep{2014Sci...344...78I}  and shape 
    \citep{2024JGRE..12908054P} provide similar values for the
    shell thickness, while the ocean below  is inferred to have a depth comparable to $d_{\rm obs}$. 
    Beneath the ocean lies a rocky   core, with radius $\sim 200$km.
     The ocean is known to be
     in contact with the  core, as deduced from the fact that particulates in 
     the water jets are  salty  \citep{2015Natur.519..207H}.

The ice shell thickness $d_{\rm obs}$ has at least two major  consequences. 
First, the librations of the shell 
distort its shape, which causes frictional heating;
the rate of energy dissipation is inversely proportional to $d_{\rm obs}$ (see eq. \ref{eq:hed} below).
Second, heat is conducted outward through the shell.  Since the temperature
jump across the shell is known, as is the conductivity of ice,  knowing the value of $d_{\rm obs}$ allows a determination
 of the moon's cooling rate.
It is found that the cooling rate is
\be
C_{\rm obs} \sim 20{\rm -}30 \  {\rm GW} \ , \label{eq:cobs} 
\ee
where 80\% of this number
comes from global conduction through an ice shell with thickness $d_{\rm obs}$.
The additional $20\%$ 
is inferred
 from  localized heat flow in the south pole region \citep{2017NatAs...1E..63L,2023SSRv..219...57N,2024JGRE..12908054P}.

The cooling rate of equation (\ref{eq:cobs}) is an important  clue.
 \cite{2007Icar..188..535M} show that Enceladus's heating
rate should be 1.1GW,  under the assumptions that the Enceladus-Dione MMR is in equilibrium, and that  $Q_S=18,000$ (in the equilibrium tide scenario).  But the fast migration
timescale found by \cite{2020NatAs...4.1053L} implies a much greater heating rate, that  exceeds even
 equation (\ref{eq:cobs}) by a factor of $\sim$4 (see eq. \ref{eq:heqnum}  below).
 Although there are large uncertainties
 in the inferred heating rate, 
if heating does not balance cooling it would
mean that the assumption
that the MMR is in equilibrium is incorrect. 
%Our proposed model provides further evidence
%that the MMR may not in equilibrium.
%which  is 
%the scenario we explore in this paper. 
% Nonetheless, the tension between heating and cooling
% could also be resolved in other ways. For example, Lainey's migration rate for
%Enceladus may be too high, or Saturn could %directly torque Dione as well as Enceladus.

\section{Overview of the Limit Cycle}
\label{sec:overview}
In this paper we  evolve Enceladus under a minimal model that includes the 
 physical processes described above: resonance locking with a mode in Saturn; 
the 2:1 MMR between Enceladus and Dione;  heat dissipation
that comes from tidal
distortions of the ice shell;
 conduction through the shell; and freezing and melting of the shell.  
We find that Enceladus naturally  settles into a limit cycle, without  artificial tweaking of
parameters.  

Before embarking on the equations and their solution, 
we provide here an overview of the limit cycle.
%%%%%%%%%%%%%%%%%%%%%%%%%%%%%%%%%%
\begin{figure}
%\centering
    %\hspace*{-.8cm} 
    \includegraphics[width=.55\textwidth,trim=10pt 0pt 0pt 0pt, clip]{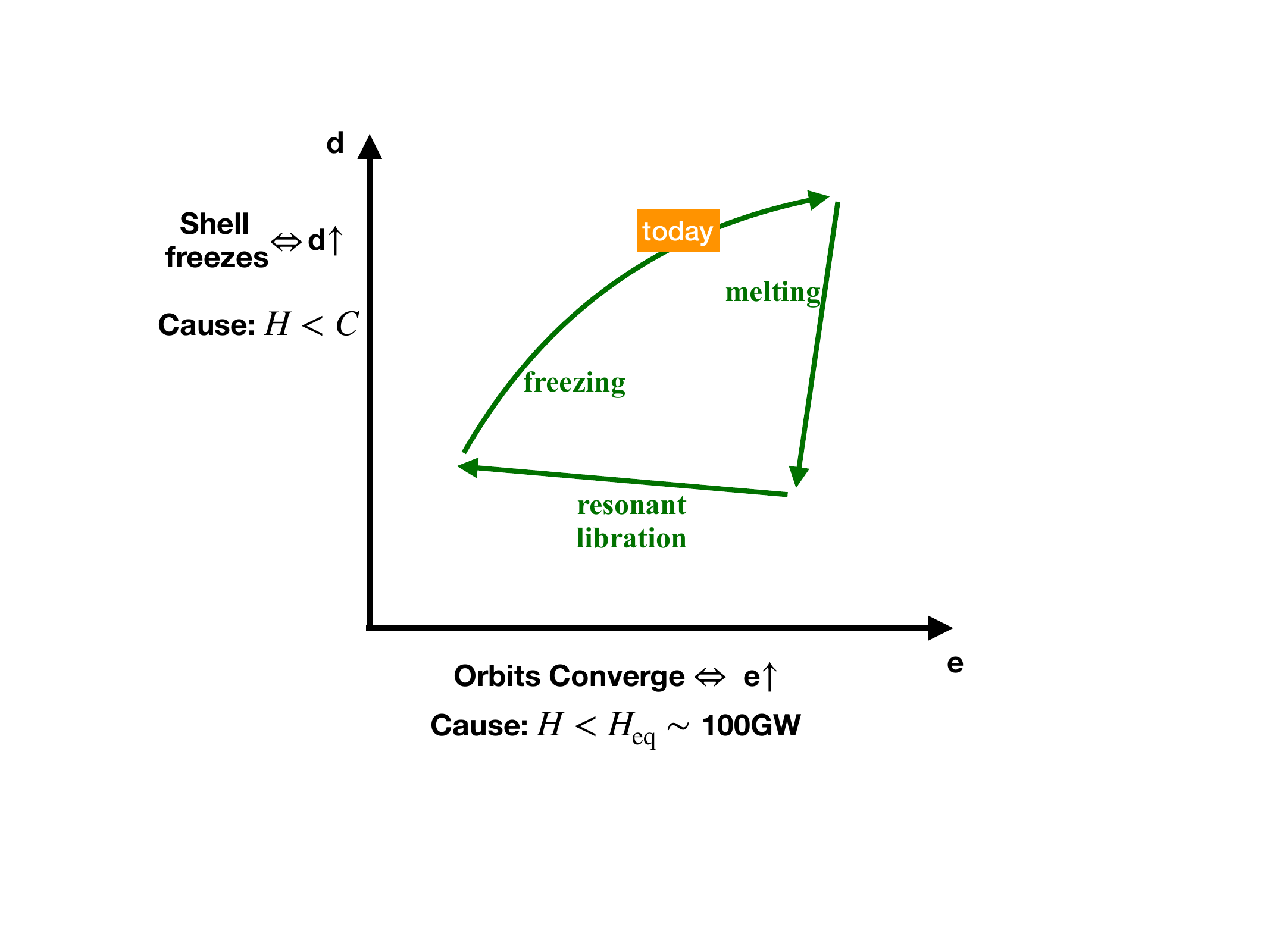}
    \vspace{-1.8cm}
\caption{{\bf Cartoon View of the Limit Cycle:} 
The axes are
Enceladus's orbital eccentricity
($e$), and
 the thickness of its ice shell ($d$) . 
}
\label{fig:cartoon}
    \vspace{1.cm} 
\end{figure}
%%%%%%%%%%%%%%%%%%%%%%%%%%%%%%%%%%
  Figure \ref{fig:cartoon}   displays a cartoon version  in the $e$-$d$ plane.
(See  Fig. \ref{fig:limitcycle} for the non-cartoon version.) 
 The $e$ axis also acts as  a surrogate
for the distance between Enceladus and Dione: $e$ rises when the moons converge, 
and falls when they diverge, due to the 2:1 MMR. 
The trajectory of the limit cycle is primarily driven by two energy rates:
 Enceladus's heating rate and its cooling rate  ($H$ and $C$). 
Heating is caused by tidal  distortions of the ice shell, and we assume that the heat
is
 deposited near the base of the  shell, where the ice
 is soft and slushy.   Cooling is due to conduction through the shell, and so
 $C\propto 1/d$.  We describe the three stages in turn, 
 starting at the freezing stage, which is where Enceladus
  finds itself today:
 
 \begin{enumerate}
 \item  Freezing:
 $d$ is increasing because $H<C$.
  To understand why $e$  also increases in the freezing stage, we  turn to the orbital dynamics, which
 set a critical heating rate at which the 
 Dione-Enceladus MMR remains in equilibrium, $H_{\rm eq}\sim 100$GW (eqs. \ref{eq:heq} \& \ref{eq:heqnum}).
 If $H>H_{\rm eq}$, the moons' orbits are driven to diverge;  otherwise, they are driven to converge. 
 The dynamics are  similar to  those of an accretion disk, where loss of orbital energy
 leads to spreading of the disk.  Here, if there is sufficient loss of orbital energy, i.e., if $H$ is large enough, 
 then the moons also diverge.
 In the freezing stage there is little heating. In particular, $H<H_{\rm eq}$, which drives  the orbits to converge, and
 $e$ increases.  The reason freezing ends is that $H\propto e^2$ (ignoring the $d$ dependence for now), because the heating is driven by
 Enceladus's epicyclic motion.
 As $e$ rises, $H$ increases, until eventually $H> C$, at which
 point freezing transitions to melting.

\item Melting. Melting is a runaway process \citep{1978Icar...36..245P}, during which $d$ decreases
by an order of magnitude, while $e$ remains nearly constant. Towards the beginning of this stage, $H\propto e^2/d$ (see eq. \ref{eq:hed} for the 
full expression), while $C\propto 1/d$.  Therefore after melting begins ($H>C$), the decrease of $d$ cannot 
tip the balance in favor of cooling. Instead, the heating rate increases, which causes more rapid melting, which
causes $H$ to increase more, leading to  runaway. 
As $d$ continues to decrease, it approaches the critical thickness for libration resonance, $d_{\rm res}\sim 3$km 
(eq. \ref{eq:dres}), and $H$ increases enormously. Shortly after $d$ sweeps through $d_{\rm res}$, 
$H$ falls sufficiently that cooling  finally becomes competitive with heating. 

\item Resonant Libration. This is perhaps the most surprising stage. The thickness of the shell remains 
 slightly below 
$d_{\rm res}$ for  millions of years, as heating and cooling nearly balance each other, 
with $H\sim C\sim 200$GW. 
The heating rate is so large because the shell's forced libration amplitude becomes very 
big near resonance, which produces large distortions. 
Almost all of Enceladus's heating occurs  during this stage. The present-day
cooling luminosity (eq. \ref{eq:cobs}) is  a modest remnant of that epoch.
An important additional consequence of the large heating is that $H>H_{\rm eq}$.  
Therefore Enceladus and Dione are pushed apart, and $e$ drops.  
This allows a new cycle of freezing to begin, with very small $e$ and hence very small heating.

\end{enumerate}

\Needspace{3\baselineskip} %prevents line break
\section{Equations of Motion}
\label{sec:eom}

\subsection{Orbital Equations}

Enceladus and Dione are observed to be in a 2:1 mean motion resonance  (MMR).
Dione is the outer moon, and lies slightly
exterior to the nominal location of the MMR.
Saturn is  pushing Enceladus outwards,
deeper into resonance with Dione.
We model Enceladus's and Dione's long-term orbital evolution 
 by evolving their angular momenta and energies via
\be
  \dot{L} + \dot{L}_2 &=& T \label{eq:torqueeq} \\
 \dot{E}+\dot{E}_2 &=& n T-  H  \label{eq:power} \ .
\ee
Unsubscripted quantities refer to Enceladus, and
ones with subscript 2 refer to Dione;
$L$, $L_2$, $E$, and $E_2$ are the angular momenta
and energies of the moons, which we express as functions of their
mean motions ($n$ and $n_2$) and eccentricities ($e$ and $e_2$), i.e., for Enceladus
\be
L&=& m(GM_S)^{2/3}n^{-1/3} \left(1-{e^2\over 2}   \right) \label{eq:lorb}
\\
E&=& -m(GM_S)^{2/3}{1\over 2}n^{2/3} \ , \label{eq:eorb}
\ee
where $M_S$ is Saturn's mass and $m$ is Enceladus's. 
The analogous equations apply to Dione's $L_2$ and $E_2$, in terms of $m_2$, $n_2$, and $e_2$. 

The quantities on the right-hand sides of equations (\ref{eq:torqueeq})--(\ref{eq:power}) are
 the torque on Enceladus by Saturn ($T$),
which also changes Enceladus's orbital energy at the rate 
$nT$ by conservation of Jacobi constant;
and $H$, which is the tidal heating rate within Enceladus.
Expressions for $T$ and $H$
are provided  below.
Although there are four unknowns ($n$, $n_2$, $e$, and $e_2$) and only two equations, 
we turn the equations into a closed set by making two further assumptions:
  that Dione's orbit is circular, and that Enceladus's 
eccentricity takes on its resonantly forced value due to the 2:1 MMR
\be
e&=& 0.76{ m_2\over M_S}{n\over n-2n_2}  \ .\label{eq:edef} 
\ee
We note that the  MMR between Enceladus and Dione has resonant argument 
$\phi = 2\lambda_2-\lambda-\pomega$, and  affects Enceladus's eccentricity, 
but not Dione's.  Dione's current eccentricity is very small (0.0022), 
and so setting it to zero is an adequate approximation.

With this setup one may   
integrate
 equations (\ref{eq:torqueeq})--(\ref{eq:power}) numerically for
 $n$ and ${n}_2$, 
once $T$ and $H$ are known.

\subsection{Torque and  Reduced Orbital Equations}
\label{sec:rl}

We model the torque $T$ within the framework
of the resonance locking mechanism \citep{2016MNRAS.458.3867F}. We assume that there is a mode in Saturn that has natural frequency
$\omega$ in the inertial frame,
 and that $\omega$ decreases on  timescale $\tau$, i.e., 
 \be 
 {\dot{\omega}\over \omega}=-{1\over \tau} \  \label{eq:omdot}
 \ee
 where $\tau$ is determined solely by the  evolution of Saturn, and  is very long ($\gtrsim$Gyr).
 Because the mode's amplitude $A$  is forced by Enceladus  it is inversely proportional to the frequency
 mismatch:
  $A\propto 1/\Delta$, 
 where 
 \be
 \Delta  &=& {\omega-n\over\omega}  \ ,
\label{eq:delta1}
\ee
which is assumed to satisfy $0<\Delta\ll 1$.
The mode loses energy in Saturn's rotating frame at a rate $\propto A^2$, which causes the moon's orbit
to gain energy at a rate $\dot{E}\propto A^2$ and, by conservation of Jacobi constant, 
to gain angular momentum at the rate $T=\dot{E}/n$.
 Consequently,
  we write $T$ in 
equations (\ref{eq:torqueeq})--(\ref{eq:power}) as
\be
T= {{\rm const.}\over \Delta^2} \label{eq:tdef} \ , 
\label{eq:trl}
\ee

 The differential equations 
 (\ref{eq:torqueeq})--(\ref{eq:power}) are time-dependent,
 via their dependence on  $T(\omega(t))$. But provided $\omega$
 decays sufficiently slowly, one may remove that dependence by changing variables
 from $n$ 
 to $\Delta$,  which is the fractional distance of
 $n$ to its ``nominal'' value of $\omega$. 
 In a similar vein,  we change
  from $n_2$ to
 \be
\Delta_2  &=& {\omega-2n_2\over\omega} \ , \label{eq:delta2}
\ee
 which  is the fractional distance of Dione's inner 2:1  MMR from its nominal value of $\omega$. 
Thus $\Delta$ and $\Delta_2$ will be our surrogates for the semimajor axes of the two moons, relative to the corotation 
radius of the resonance locking mode.
The nominal MMR  is at $\Delta=\Delta_2$, and we will always have $\Delta<\Delta_2$, i.e., Dione is exterior to nominal MMR. 

 As we show in  Appendix \ref{sec:appendix}, 
 the resulting equations for $\Delta$ and $\Delta_2$, under the assumption that $\Delta\ll 1$ and $\Delta_2\ll 1$, 
 are
\be
 {\epsilon}\dot{\Delta}+\dot{\Delta}_2
 -3\epsilon e\dot{e}   &=&
\left({\Delta_{\rm eq}^2\over \Delta^2}-1\right){1+\epsilon \over \tau}  \label{eq:torque_2}
 \\
\dot{\Delta}_2 -6\epsilon e\dot{e}&=&
\left( {H\over H_{\rm eq}}-1  \right){1\over \tau} \ ,
\label{eq:power_2}
\ee
where
\be
\epsilon\equiv {m\over 2^{1/3} m_2 } \label{eq:eps}
\ee
 is the ratio of the  moons' nominal angular momenta, 
 $\Delta_{\rm eq}$
 is a constant that replaces the unknown constant
 in equation (\ref{eq:tdef}), 
 and
 \be
 H_{\rm eq} 
 &=& {m_2\left(GM_Sn  \right)^{2/3}\over  3\cdot 2^{2/3}\tau}
 \label{eq:heq}
 \ee
  is the equilibrium heating rate,\footnote{
Equation  (\ref{eq:heq}) 
is the heating  rate when the MMR is in equilibrium; 
i.e., it follows
   from equations (\ref{eq:torqueeq})--(\ref{eq:power}) after setting
$n=2n_2$, 
  $\dot{n}/n=-1/\tau$, and neglecting both $e^2$ and Saturn's direct torque on Dione, as shown by \cite{2007Icar..188..535M} in the context of the equilibrium
 tide scenario. 
 }
   with the symbol $n$ now considered   to be constant.
   Equations (\ref{eq:torque_2})--(\ref{eq:power_2})
are to be supplemented with
 \be 
 e =  0.76 {m_2\over M_S}{1 \over \Delta_2-\Delta} \  , \label{eq:enum}
 \ee
 and an expression for $H$ (provided below),
 whereupon they
 form a closed set of differential equations for $\Delta$ and $\Delta_2$, 
without  explicit time-dependence. 

When evaluating expressions, we set 
\be
m&=&1.08\times 10^{23} {\rm g} 
\\
 m_2&=&1.095\times 10^{24} {\rm g} 
 \\ 
 M_S&=&5.683\times 10^{29} {\rm g} 
 \\
 n&=&2\pi/(1.3702{\rm day}) 
 \ee
 whence
 \be
 \epsilon&=&0.078 \\
H_{\rm eq} &=& 116 {\rm GW}\left({10 {\rm Gyr}\over \tau} \right)  \label{eq:heqnum} 
\ee

\subsection{Tidal Heating}
\label{sec:heating}

%%%%%%%%%%%%%%%%%%%%%%%%%%%%%%%%%%
\begin{figure*}[t!]
\centering
    \begin{minipage}[b]{.48\textwidth}
    \centering
    \includegraphics[width=\textwidth]{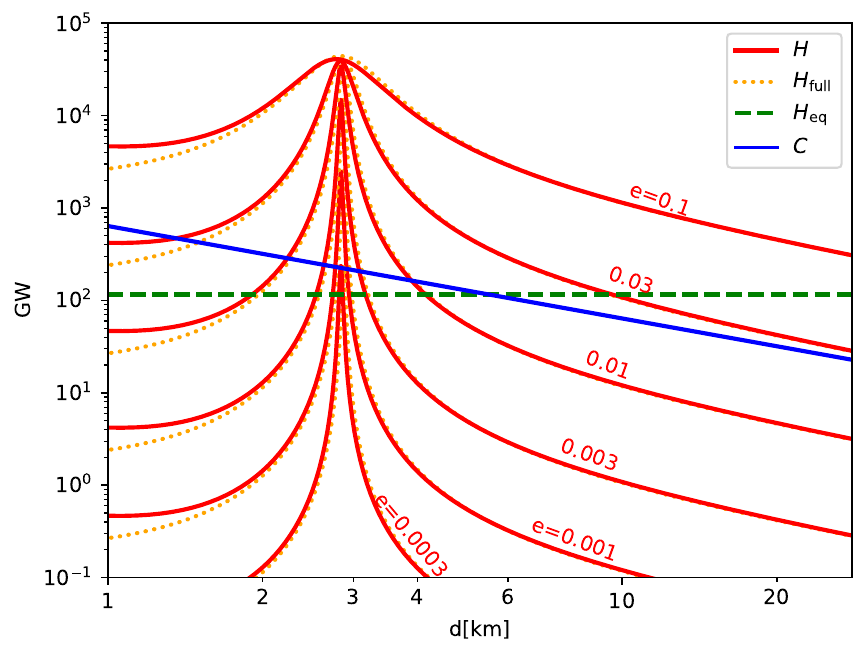}
    \end{minipage}
%\hfill
    \begin{minipage}[b]{.47\textwidth}
    \centering
    \includegraphics[width=\textwidth]{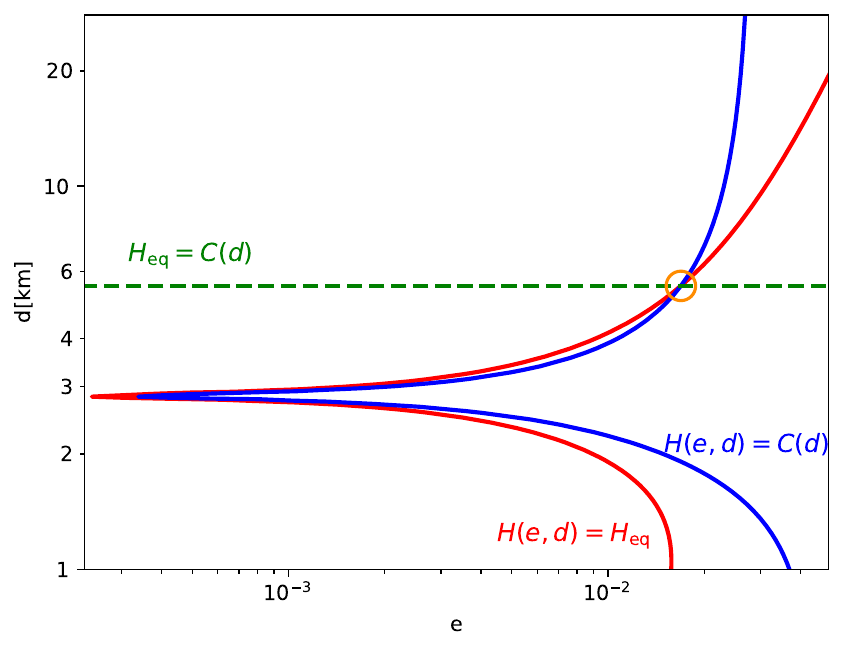}
    \end{minipage}
\caption{ Heating and Cooling Rates. {\bf Left panel:}
 The red curves are the tidal heating rate, $H(e,d)$, at selected values of $e$, 
setting $\Q=25$.
The orange dotted curves  ($H_{\rm full}$) are for a more general heating model that does not assume that
the shell is very rigid. 
 But since $H_{\rm full}$ leads to little change
over the range of $d$'s of interest, we use 
$H$
in this paper for simplicity.
The green horizontal dashed line is the
 heating rate  for equilibrium pushing (eq. \ref{eq:heq}, at
 our fiducial value of $\tau=10$Gyr).
And blue is the cooling   due to conduction through the shell (eq. \ref{eq:cool}). 
{\bf Right panel:} Equilibrium curves in the $e$-$d$ plane,  produced by equating   rates in the left panel.  The red curve represents equilibrium of the MMR, and the blue curve is where the ice shell thickness does not change. The  circle is the global (unstable) equilibrium point.}
\label{fig:heating}
\vspace{1cm} 
\end{figure*}
%%%%%%%%%%%%%%%%%%%%%%%%%%%%%%%%%%

Tidal heating within Enceladus ($H$) is assumed to be due to  deformations of its ice shell.
We model Enceladus as being composed of a   rigid solid
core  covered by an ocean, 
which in turn is covered by a thin ice shell.
The undistorted core and shell are taken to be spherically
symmetric. 
If Enceladus were on a circular orbit it would
be in a spin-synchronous state, and the shell's
   shape would be determined by the lowest energy 
response to Saturn's tidal field. 
In that case, there would be no energy dissipation.
But because $e\ne 0$,  time-dependent forcing from 
Saturn distorts the shell, which dissipates energy.
We calculate
the time averaged elastic energy within the ice shell, $\langle E_{\rm elas}\rangle$, that is induced
by the epicyclic motion of Saturn, as seen from Enceladus. We then set 
 the
heating rate to $H={2n\over \Q}\langle E_{\rm elas}\rangle$, where $\Q$ is the quality factor 
for the elastic flexure of the shell, and is set to a constant value. 
See the companion paper Lithwick (2025) for the full
calculation. 
That calculation
 is similar to \cite{2013Icar..226..299V} in physical
 content, but yields an  analytic
 expression for the  heating rate.
 %which we  call here $H_{\rm full}$.
 In Figure \ref{fig:heating} (left panel), we plot the
 resulting heating rate 
$H$ vs. shell thickness ($d$), for various values of $e$, 
as solid red curves. 
The heating spikes at
 $d_{\rm res}=2.8$km are due to
%\be
%d_{\rm res} &\approx& {9\over 10}R{n^2R\over g}h_* \\
%&=&
%2.8 {\rm km} \ , 
%\ee
%where $R=252$km is Enceladus's radius, 
%$g=Gm/R^2$ is its surface gravity, and the %dimensionless number $h_*\approx 2$.
%At thickness $d_{\rm res}$ 
 a libration  resonance. 
 In particular, when the shell's thickness
 is equal to $d_{\rm res}$, 
the frequency of the shell's free librations is equal 
to the orbital frequency.  That resonance drives the
shell to  a large forced libration amplitude, which
is the cause for the enhanced heating. 
The enhanced heating at the libration resonance will play an important role
in the limit cycle behavior. 

The analytic expression for the heating curves in the figure
is as follows (Lithwick 2025):
 \interdisplaylinepenalty=10000
\be
H(e,d) &=&{21\pi\over 5}{\rho_w^2 R^8 n^5\over \mu  \Q} 
{e^2\over  d} \times \nonumber 
\\ &&
\left({3\over 7}   +	   {4\over 7}{1\over 
	   \left({\omega_{\rm lib}^2/ n^2}-1\right)^2+ \left(\Q\R \right)^{-2} + 
	   4e^2
	  }
	  \right)  \ \ \ \  \ 
	  \label{eq:hed}
\ee
\interdisplaylinepenalty=100
where
$R=252$km is Enceladus's radius; 
 $\rho_w=0.93$gm/cm$^3$ is the density of water, which we take here to be the same as that  of ice; $\mu=4$ GPa is the rigidity of   ice; and  
 the Poisson ratio of  ice has been set to  $1/3$.  The  first term in brackets (3/7) is from the radial 
tide, and the second is from the librational tide \citep{1999ssd..book.....M}.  The resonant denominator 
in the librational tide depends primarily on the frequency of free librations of the ice shell, 
\be
{\omega_{\rm lib}^2\over n^2} &=& {d_{\rm res}\over d} 
\ee
where
\be
d_{\rm res}&=&{9\over 10}R{n^2R\over g}h_*   
\\
&=&  2.8{\rm km} \label{eq:dres} ,
\ee
 $g=Gm/R^2$, and the dimensionless number $h_*\approx 2$, which
 incorporates the effect of a rigid core (Lithwick 2025).
Finally, the ``hardness parameter'' is 
$\R\approx d$/(0.84 km) for Enceladus. 

In general, the hardness parameter can play an important role
\citep{2010Icar..209..631G}.  But
for Enceladus $\R\gtrsim 1$ throughout the evolution.  
Physically, the limit $\R\gg 1$ corresponds to the shell
being very rigid, which means that it 
hardly changes its shape as
it undergoes its forced librations.
In writing equation 
(\ref{eq:hed}), 
 we have therefore chosen to simplify it, for clarity, 
 by taking the limit
$\R\gg 1$. That is why $\R$ only plays a modest role in that equation.
But we  show 
in the figure  what happens for arbitrary values of $\R$,  as orange dotted curves ($H_{\rm full}$).  
%See Lithwick (2024) for the full expression. 
 The match between solid
and dotted curves is adequate for $d\gtrsim 1.5$km. Hence we adopt the above simplified expression for
$H$ for the remainder of this paper.

\subsection{Freezing and Melting}

The thickness of the ice shell, $d$, affects the evolution via its effect on $H(e,d)$.
And $d$ is in turn affected by tidal heating, which can melt the ice. 
To model the evolution of $d$, 
 we assume that most of the heating is deposited near the
base of the ice shell, because the slushy mixture there is much 
more dissipative than the solidly frozen ice higher up.
The rate at which energy leaves the moon (the cooling rate) is
\be
C(d)&=& k_{\rm ice}{\Delta T\over d}4\pi R^2 \label{eq:cool} % \\
\ee
where
$k_{\rm ice}=4\times 10^5$erg/cm/K is the
conductivity of the shell, which is approximated to be constant,  
and $\Delta T=200$K is the temperature jump across the shell.
We  set the net heating
equal to the rate of energy change due to 
freezing/melting
i.e.,
\be
H(e,d)-C(d)= -4\pi R^2\rho_w\left(\ell+c_p{\Delta T\over 2}   \right)\dot{d}
\label{eq:melt}
\label{eq:ddot}
\ee
where 
$\ell=3.34\times 10^9$erg/g is the latent heat of fusion, and $c_p= 10^7$erg/g/K is the heat capacity of ice. Note that
the second term in brackets is due to the change in internal energy, as melting of ice  necessitates that the remaining ice warms up.
Equation (\ref{eq:ddot}) 
is our final evolutionary 
equation; it gives $\dot{d}$ as a function
of the variables $e$ and $d$.
In the absence of heating ($H(e,d)\rightarrow 0$), our adopted
parameters imply a freezing timescale $d/\dot{d}= 6.4$Myr 
$\times \left(d/(20 {\rm km})\right)^2$.
%Combining the above equations produces our third and final evolutionary equation:
%\be
%\dot{d} &=& {C(d)-H(e,d)\over 4\pi R^2\rho_w\left(l+c_p{\Delta T\over 2}  \right)}  \ee

Equation (\ref{eq:cool})  is  shown in the left panel of Figure \ref{fig:heating} for our fiducial parameters. 
As described in \S \ref{sec:overview}, the limit cycle is largely controlled by the relative values
of $H$, $C$, and $H_{\rm eq}$.  Therefore in the right panel of Figure \ref{fig:heating}, we display
the three resulting equilibrium curves in the $e$-$d$ plane.

 \subsection{Summary of  Equations, Choice of Parameters, and Global Equilibrium}
 \label{sec:sum}
 
 The equations of motion for $\Delta$ and $\Delta_2$,  which encode the locations of the
 two moons, are given by
 equations (\ref{eq:torque_2})--(\ref{eq:power_2}), 
 supplemented with equations (\ref{eq:eps})--(\ref{eq:enum}).
 For the heating rate $H$, we use 
equation (\ref{eq:hed}). 
 The equilibrium of equations  (\ref{eq:torque_2})--(\ref{eq:power_2})  has
  both moons  migrating in lockstep with  the frequency of the driving mode within Saturn.
 Zeroing the time derivatives in those equations  gives the equilibrium state:
 \be
 \Delta&=&\Delta_{\rm eq} \\
H(e,d)&=&H_{\rm eq} \label{eq:eq1} \ ,
 \ee
 where the latter is the red curve in the right panel of Figure \ref{fig:heating}.
 The third and final equation of motion is for $d$. It is given by equation (\ref{eq:ddot}), for which $H(e,d)$ is also needed.   
Its equilibrium occurs when
\be
H(e,d) &=& C(d)  \ , \label{eq:eq2}
\ee
which is the blue curve in the right panel of Figure \ref{fig:heating}.

 For the bulk of this paper, we   choose the following   values for the three  uncertain parameters
 in our model:
 \be
\tau&=& 10\ {\rm Gyr} \\
\Q&=& 25 \\
\Delta_{\rm eq}&=&  0.02  \label{eq:delta1eq}
\ee
The value of $\tau$ is comparable to the one inferred by \cite{2020NatAs...4.1053L}.
For $\Q$, we expect it to be much larger than unity, because most of the ice shell
is solidly frozen. 
And  although the value of $\Delta_{\rm eq}$ is 
very uncertain,  we find 
it does not play a significant role.  
Towards the end of the paper we show what happens
when these parameters are varied.

We conclude this subsection by solving for the global equilibrium, $H=C=H_{\rm eq}$, 
which will be used
 as the
initial condition of the integration.
The
 equilibrium thickness of the ice shell follows
 from equating ${\cal H}_{\rm eq}=C(d)$, which
gives
$d_{\rm eq}= 5.50$ km. 
This value is proportional to $\tau$, but is 
 uninfluenced by  other uncertain parameters.  
To obtain the equilibrium $e$, we insert
$d_{\rm eq}$ into equation (\ref{eq:eq1}), from which we obtain numerically
$e_{\rm eq} =  0.017$.
One  then finds via equation (\ref{eq:enum}) that 
$\Delta_{2,\rm eq}=\Delta_{\rm eq}+8.7\times 10^{-5}$.

\subsection{Numerical Integration}
In order to  integrate the equations of motion,
we use equation (\ref{eq:enum}) to
 replace $e\dot{e}\rightarrow -{e^2\over\Delta_2-\Delta}(\dot{\Delta}_2-\dot{\Delta})$ in
equations (\ref{eq:torque_2})--(\ref{eq:power_2}), and  at each timestep we solve
the latter two equations
algebraically
for  $\dot{\Delta}$ and $\dot{\Delta}_2$.
We then
integrate those time-derivatives, along with equation (\ref{eq:ddot}) for $\dot{d}$.

\section{Evolution}

\subsection{From Unstable Equilibrium to Limit Cycle}

%%%%%%%%%%%%%%%%%%%%%%%%%%%%%%%%%%
\begin{figure*}[t]%[ht!]
\centering
    \begin{minipage}{.48\textwidth}
    \centering
    \includegraphics[width=\textwidth]{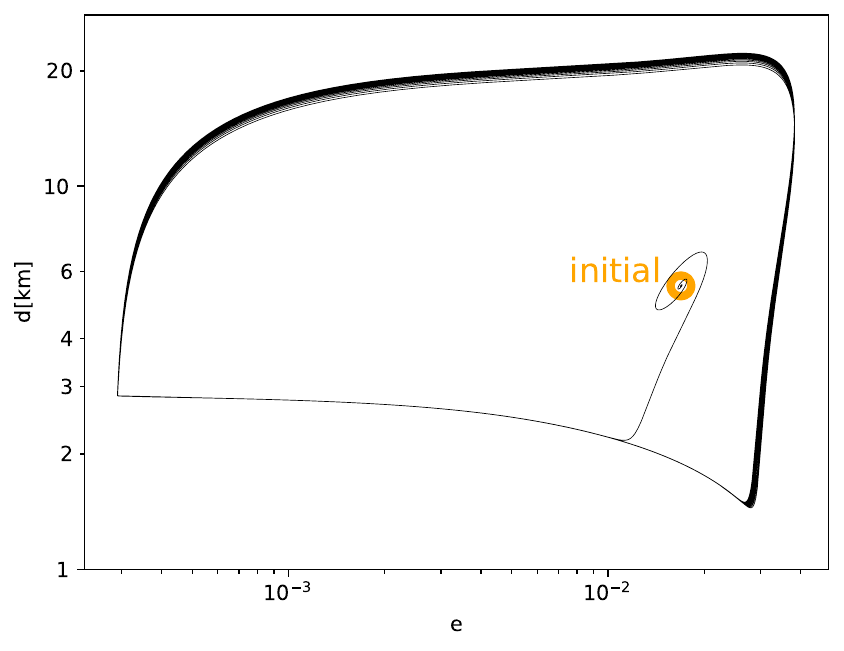}
    \end{minipage}
    \begin{minipage}{.48\textwidth}
    \centering
    \includegraphics[width=\textwidth]{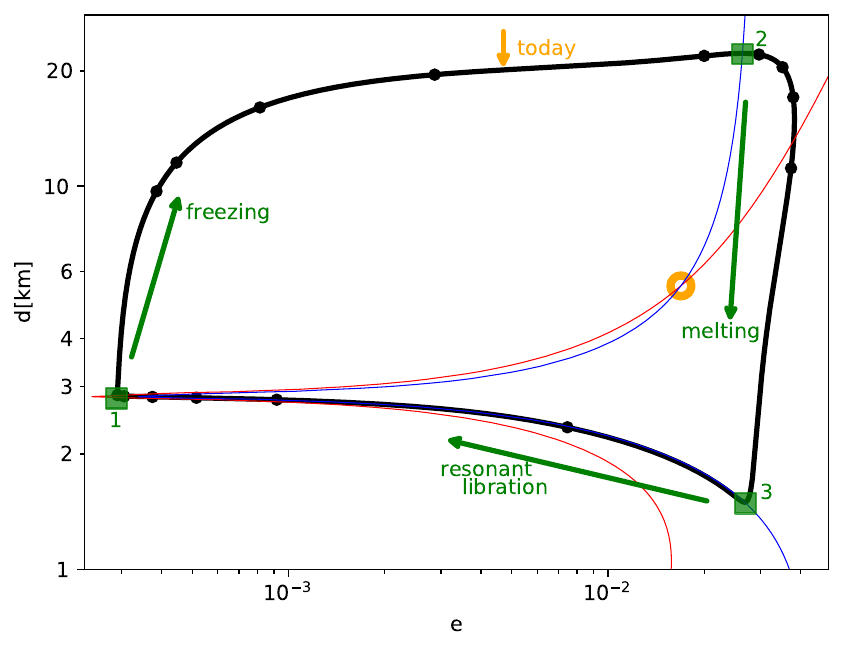}     
    \end{minipage}
\caption{Numerical Evolution in the $e$-$d$ 
Plane.  {\bf Left panel:} 
The black curve displays the system's evolution, after it is
  initialized at the global equilibrium point. It  escapes from equilibrium, 
and then traces out repeating limit cycles.
{\bf Right panel:} The black curve is the same as in the left panel, but restricted to the limit cycle. 
The black circles on the curve are separated by 1 Myr, over the course of one cycle.  The blue and red  curves are the equilibrium heating and cooling curves, repeated from Figure \ref{fig:heating},
with the blue showing $H=C$, and the red showing $H=H_{\rm eq}$.
.
}
\label{fig:limitcycle}
\vspace{1cm} 
\end{figure*}
%%%%%%%%%%%%%%%%%%%%%%%%%%%%%%%%%%

We initialize the integration at the global equilibrium point 
presented at the end of \S \ref{sec:sum}. 
In the $e$-$d$ plane, the system spirals away from its initial state, and  
 converges to a repeating
 limit cycle, as shown
 in the left panel of Figure \ref{fig:limitcycle}.
 Each cycle lasts 
 13.3 Myr.
The right panel
shows   the steady cycle, with
  the freezing, melting, and resonant libration stages  marked out:
\bi
  \item Freezing (1 $\rightarrow$ 2): The shell thickens from $2.8$km to $22$km, while Enceladus's eccentricity rises because it is being pushed deeper into its MMR with Dione.  
  Enceladus is currently in this phase. The vertical orange arrow marks where $e=e_{\rm obs}=0.0047$.
  \item Melting (2 $\rightarrow$ 3): The shell melts from $22$km to $1.5$km, while Enceladus's eccentricity remains high, at $e\sim 0.04$.
  \item Resonant Libration (3 $\rightarrow$ 1):  the shell's thickness  slowly  grows towards $d_{\rm res}=2.8$ km, the thickness for exact resonant libration  (eq. \ref{eq:dres}).  Most of the heating  occurs during this phase, and one consequence of the extreme heating is that
  Enceladus is driven away  from Dione's MMR, lowering its resonantly forced
  eccentricity to  0.0003.
\ei

The black circles in this panel are separated by 1 Myr. 
Also shown are the equilibrium curves for heating and cooling, repeated from the right panel of Figure \ref{fig:heating}.
At point 2, melting begins when the $H=C$  curve is crossed. The subsequent melting phase 
 is seen to be a runaway process: while  millions of years are required to melt the shell thickness by a factor of $\sim 2$, the shell then melts to 1.5km in under a million years.
Immediately before Enceladus crosses  point 3, its cooling rate grows to around  $10 C_{\rm obs}$, because
 $C\propto 1/d$.   Even more dramatically, the heating rate spikes to $\sim 1000C_{\rm obs}$, 
 because $d\sim d_{\rm res}$. 
Immediately after point 3 heating is very high, albeit much less than in the aforementioned spike. 
 That  forces $e$ to drop very quickly
  to $\sim 0.005$, by pushing the moons apart.  
 Subsequently, the  evolution in the resonant libration stage slows down, as the boost in heating caused
 by $d\sim d_{\rm res}$ is counterbalanced by the decrease in heating due to a very small $e$. 
Millions of years
 elapse   at the end of the resonant libration stage, before the system  re-enters  its current freezing phase.

%%%%%%%%%%%%%%%%%%%%%%%%%%%%%%%%%%
\begin{figure*}
\centering
    \begin{minipage}[b]{.45\textwidth}
    \centering
    \includegraphics[width=\textwidth]{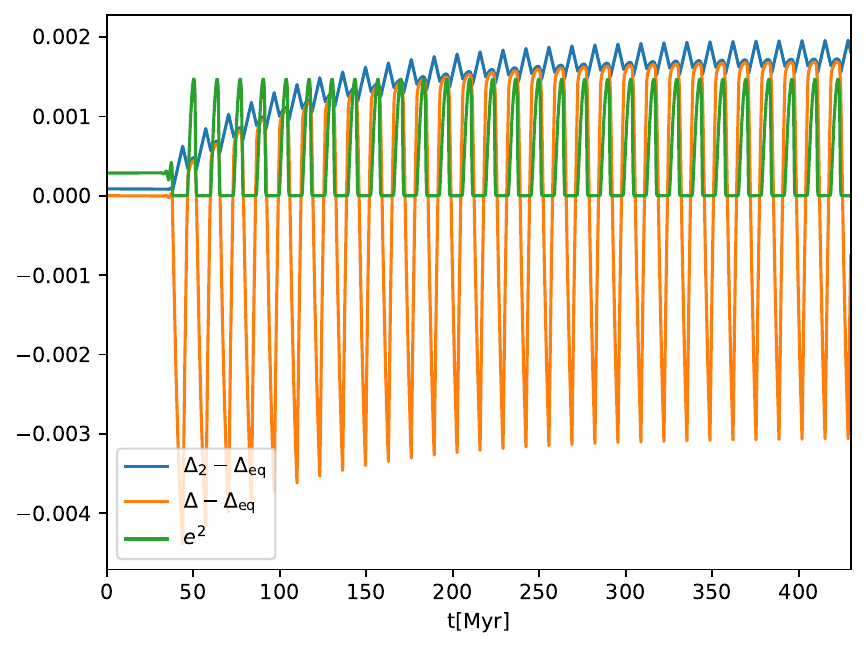}
    \end{minipage}
    \begin{minipage}[b]{.45\textwidth}
    \centering
    \includegraphics[width=\textwidth]{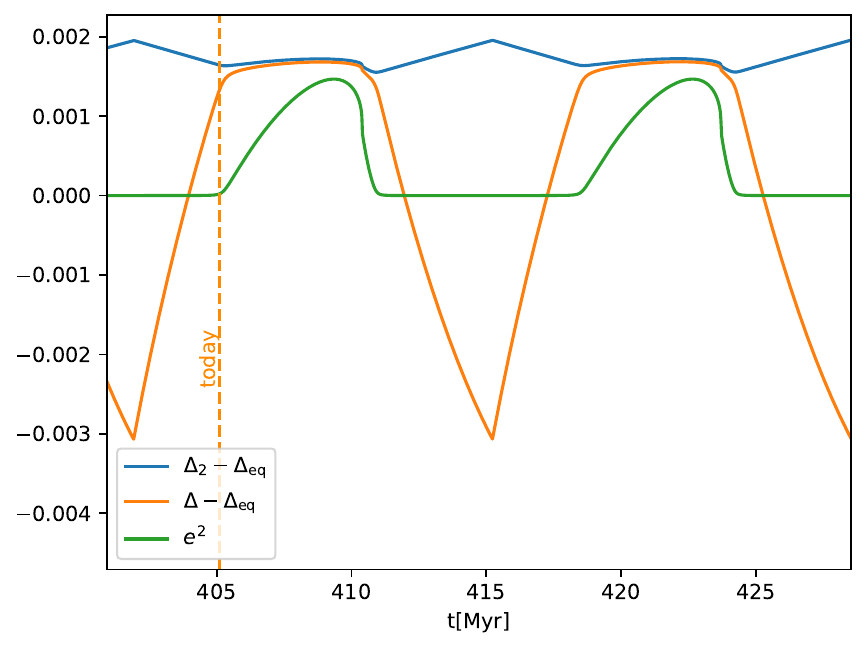}
    \end{minipage}
\caption{Evolution of $\Delta$, $\Delta_2$, and $e^2$, where
we have subtracted the constant
 $\Delta_{\rm eq}=0.02$ from the $\Delta$'s. 
{\bf Left panel:}
Initial evolution.
{\bf Right panel:}  Evolution in the limit cycle.
The vertical line marked ``today'' shows when $e=e_{\rm obs}=0.0047$. 
}
\label{fig:delta}
\vspace{1cm} 
\end{figure*}
%%%%%%%%%%%%%%%%%%%%%%%%%%%%%%%%%%
Figure \ref{fig:delta} highlights the role of Dione's MMR in the evolution by showing    $\Delta$ and $\Delta_2$ at early times (left panel) and in the steady limit cycle (right panel).
Recall that
 $\Delta$ is the fractional distance, in frequency-space,  of Enceladus from its  nominal position $\omega$, 
 and $\Delta_2$  is the fractional distance of Dione's MMR from $\omega$ (eqs. \ref{eq:delta1}\&\ref{eq:delta2}).
From the left panel, we see that it takes over 100Myr for the moons' positions to reach their final limit cycle. 

\subsection{Behavior in the  Limit Cycle}

In the right  panel of Figure \ref{fig:delta}, the first minimum of the orange curve at 402Myr corresponds
 to the beginning of the freezing stage (point 1 in right panel of Fig. \ref{fig:limitcycle}). Following that time, the 
 orange curve increases, indicating outward motion of Enceladus, and the blue
 curve decreases, indicating inward motion of Dione. 
 The moons continue to converge
 until $e^2$ (green curve) hits its maximum, at 409Myr.
 Note 
  that $e$ 
 is inversely proportional to
 $\Delta_2-\Delta$,
    and so it tracks convergence/divergence. 
  In the limit cycle plot (right panel of Fig. \ref{fig:limitcycle}), the maximum $e$ occurs
  when the red $H=H_{\rm eq}$  curve is crossed, because convergence is driven
  by $H<H_{\rm eq}$, and ends when $H=H_{\rm eq}$.

   The moons converge from 402 to 409Myr. But there is a qualitative transition
    at $\sim$405Myr. Before that time,
    there is rapid convergence
   of the moons' $\Delta$'s, or equivalently of their semimajor axes, while $e$ remains very low. 
   But the subsequent convergence is much slower, as the moons are already very close to exact resonance, 
   and further convergence drives the rise of
   $e$.
    Those two types of behavior are evident from the equation of motion (eq. \ref{eq:power_2}); each type corresponds
    to one of the two terms on the left-hand side
    dominating the other. 
   After 409Myr, the moons' orbits diverge, until the next cycle begins.  Virtually all of the divergence occurs during the resonant
   libration stage.

%%%%%%%%%%%%%%%%%%%%%%%%%%%%%%%%%%
\begin{figure}
    \includegraphics[width=.4\textwidth]{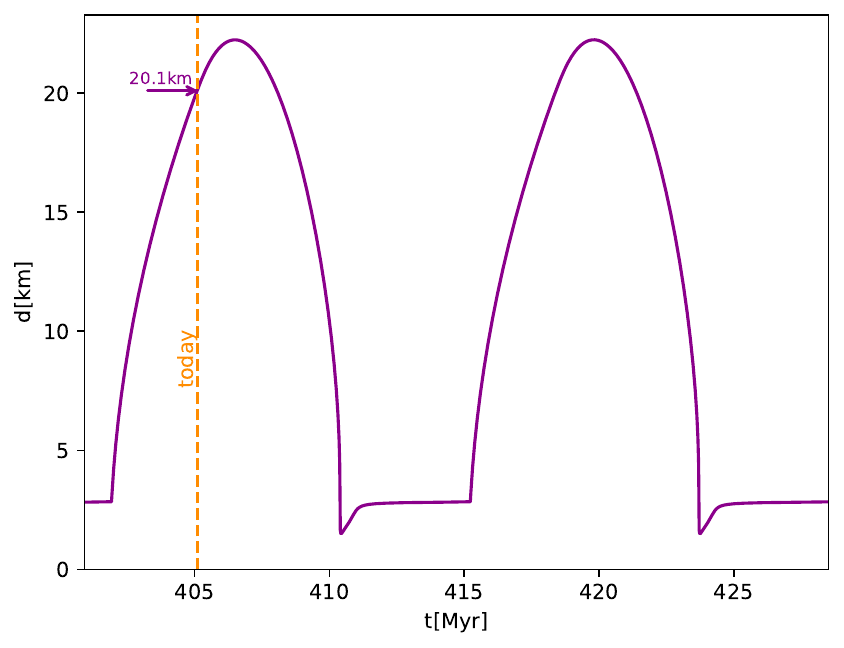}
    \includegraphics[width=.4\textwidth]{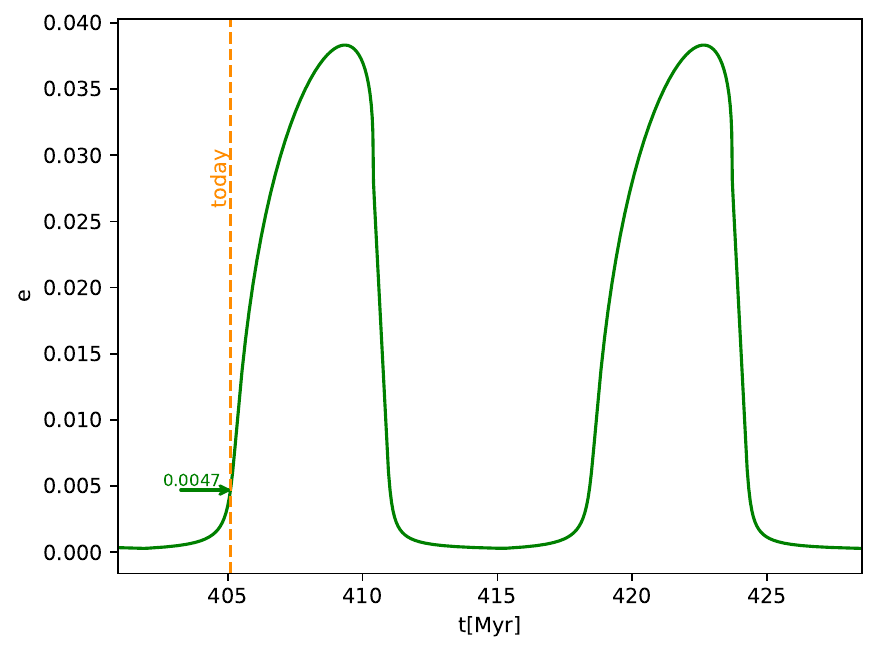}
    \caption{Evolution of 
$d$  and $e$ over two limit cycles.
}
\label{fig:de}
\vspace{1cm} 
\end{figure}
%%%%%%%%%%%%%%%%%%%%%%%%%%%%%%%%%%
Figure \ref{fig:de} shows the temporal evolution of $d$ and $e$
 over the course of two limit cycles. 
The vertical line  (``today'') is when $e=e_{\rm obs}=0.0047$ and $d$ is growing. Of course, 
that happens once per cycle, but we arbitrarily choose to focus on the cycle beginning at 402Myr. 
At the current time, $e$ has just commenced  rapid growth, while $d$ is already close to  its maximum
value.  
The   shell thickness 
is currently 20.1km in the model, which is comparable to $d_{\rm obs}$ (eq. \ref{eq:dobs}).
In the resonant libration stage, most of the time is spent with $d$ and $e$ fairly constant, 
before that stage ends at $415$Myr. The enormous heating and cooling that occur at the
start of this stage are not very apparent in  this figure.  They are associated with 
the small rise in $d$ immediately after 410.4Myr, and the concurrent extremely rapid decline in $e$.

%%%%%%%%%%%%%%%%%%%%%%%%%%%%%%%%%%
\begin{figure}[h]
\centering
    \begin{minipage}[b]{.4\textwidth}
    \centering
    \includegraphics[width=\textwidth]{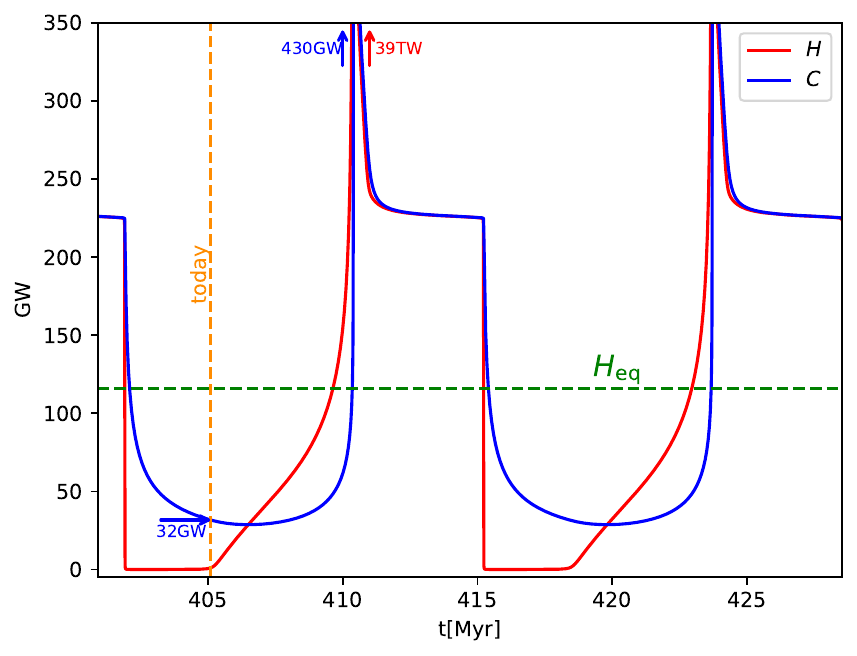}
    \end{minipage}
\hfill
    \begin{minipage}[b]{.4\textwidth}
    \centering
    \includegraphics[width=\textwidth]{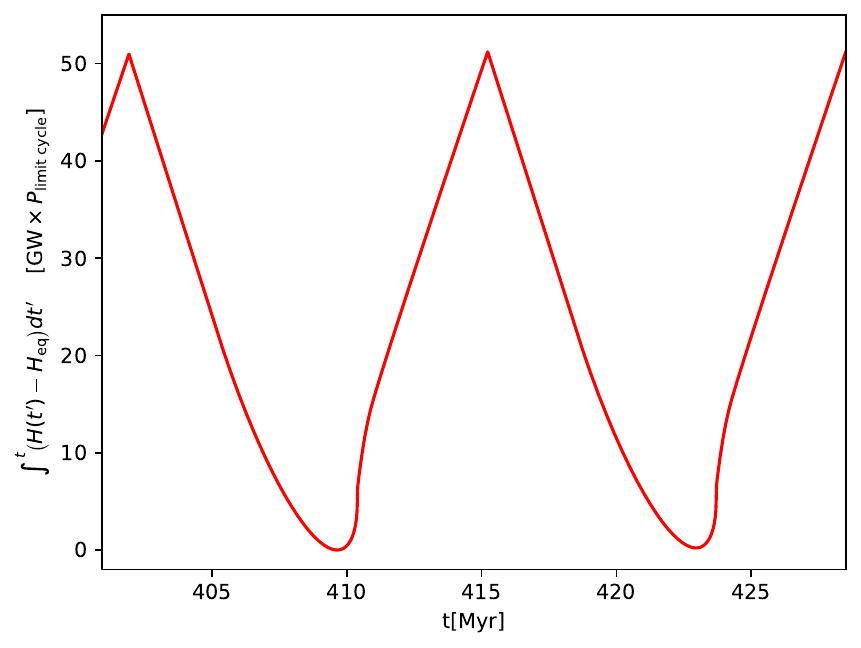}
    \end{minipage}
\caption{{\bf Top panel:} Heating and Cooling Rates. 
{\bf Bottom panel:} Integrated heating rate.
}
\label{fig:heat}
\vspace{1cm} 
\end{figure}
%%%%%%%%%%%%%%%%%%%%%%%%%%%%%%%%%%
Figure \ref{fig:heat} (top panel) shows the heating and cooling rates, 
which drive most of the system's behavior. At the current time 
the cooling rate is $C=32$GW, consistent with the observationally inferred $C_{\rm obs}$  (eq. \ref{eq:cobs}).
The heating rate is much smaller than that,  $H=1$ GW.
The current value of $H$ is so small primarily because $e$ is small. But  the two moons are presently converging, 
and $e$ is about to rise dramatically (right panel of Fig. \ref{fig:delta}). Thus, after 405Myr, $H\propto e^2$ also rises
dramatically, until $H$ crosses $C$.\footnote{
Although
 $H$   depends on the shell thickness $d$ in addition  to $e$, with $H\propto e^2/d$, 
 both $H$ and $C$  are  inversely proportional to $d$, and so the $d$-dependence does not 
 affect the relative magnitude of the two rates; moreover, the rise
 in $e$ affects $H$ much more than does the rise in $d$ at this time.}  The time when $H=C$, which occurs 
 at 406.5Myr, marks the beginning of the melting stage. In the limit cycle plot 
 (Fig. \ref{fig:limitcycle}), that time is labelled 2, which is when the
 blue curve is crossed. 
Returning to the top panel of Figure \ref{fig:heat}, we see that $H$ continues to rise past $C$, until 
it reaches $H=H_{\rm eq}$ at 409Myr.  As noted previously, that marks the time that the moons stop 
converging, and begin to diverge. Divergence lowers $e$ and reduces $H$; but the timescale for that
to happen is relatively long.  Instead, runaway melting occurs first: with $H>C$, and both of those
rates nearly inversely proportional to $d$, the shell  melts at the rate $\dot{d}\approx  -{\rm const}/d$ (eq. \ref{eq:melt}).
If that equation continued to apply, the shell would melt to $d=0$ in a finite time. 
In actuality, the shell does initially get thinner at an increasingly rapid rate (Fig. \ref{fig:de}, top panel). 
But before it hits $d=0$, it crosses the critical value for resonant libration, $d=2.8$km, when the heating
rate becomes enormous: $H= 39$TW, as seen from the heating curves  of Figure \ref{fig:heating} (left panel). 
As $d$ continues to decrease beyond the point where $\omega_{\rm lib}=n$, the librational tide
acquires a dependence on $d$ (eq. \ref{eq:hed}), which allows cooling to finally catch up with heating, thereby 
preventing the shell from melting entirely. 
The shell melts to a thickness of 1.5km, whereupon the resonant libration stage commences, 
at 410.4Myr. In the limit cycle plot (Fig. \ref{fig:limitcycle}), that occurs at point 3, 
where the blue $H=C$ curve is crossed again. 
The enormous spike in  $H$ immediately before the end of the melting stage  affects the energy budget.
But because it lasts a short time, the net effect is modest. 
From  the bottom panel of
Figure \ref{fig:heat}, which shows the integrated heating rate, the spike immediately before 410.4Myr injects an extra energy of
 6.5GW$\times P_{\rm limit\ cycle}$ 
 (on top of what would be injected if
 $H=H_{\rm eq}$), where $P_{\rm limit\ cycle}=13.3$Myr
 is the period of a limit cycle. In comparison, the subsequent resonant libration stage injects
 an extra energy of 44.8GW$\times P_{\rm limit\ cycle}$.
 Thus the resonant libration stage is primarily responsible for heating Enceladus. 
 That conclusion is also apparent from the limit cycle plot (Fig. \ref{fig:limitcycle}), which shows that most of the
 drop in $e$---which is driven by $H-H_{\rm eq}$---occurs during resonant libration. 
%%%%%%%%%%%%%%%%%%%%%%%%%%%%%%%%%%
\begin{figure}[h]
\centering
    \begin{minipage}[b]{.4\textwidth}
    \centering
    \includegraphics[width=\textwidth]{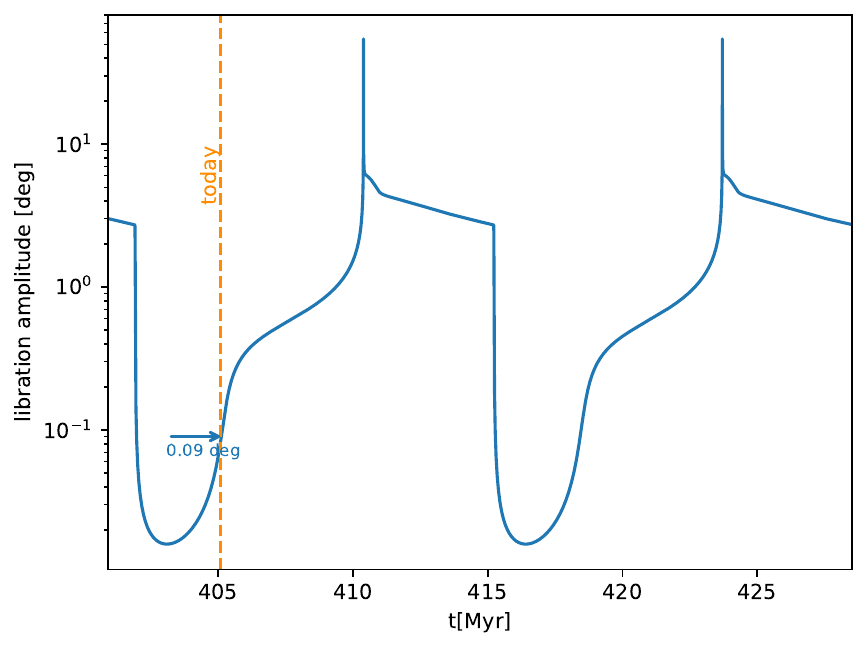}
    \end{minipage}
\caption{Libration Amplitude
}
\label{fig:lib}
\vspace{1cm} 
\end{figure}
%%%%%%%%%%%%%%%%%%%%%%%%%%%%%%%%%%
Figure \ref{fig:lib} shows the shell's libration amplitude $\hat{\gamma}_{\rm ice}$, which is approximately given by
\be
\hat{\gamma}_{\rm ice} \approx 
2 e 
{\omega_{\rm lib}^2\over |\omega_{\rm lib}^2-n^2|
}
\ee
(See Lithwick 2025 for the full expression that is plotted in the figure.)
At the current time, $\hat{\gamma}_{\rm ice}=0.09^\circ$, in agreement with the observed $\gamma_{\rm obs}$ (eq. \ref{eq:gammaobs}).  And during the extreme spike in heating at the end of the melting stage, $\hat{\gamma}_{\rm ice}$ reaches 
as high as $50^\circ$, i.e., the shell rotates nearly a full revolution each orbital period.\footnote{The narrow spikes in 
the libration amplitude (Fig. \ref{fig:lib}) are sufficiently broad that the shell executes many librations over the course of the spike, as is required for the heating rate formula  (eq. \ref{eq:hed}) to remain applicable. Quantitatively, the fastest 
rate of change of the libration amplitude 
is $0.3^\circ$/yr, which implies a small change in 
amplitude  over the course of a single libration period  (1.3702 days). }
Of course, it is that large libration amplitude
that drives the large heating rate.

%%%%%%%%%%%%%%%%%%%%%%%%%%%%%%%%%%
\begin{figure*}[t!]
\centering
    \includegraphics[width=1\textwidth]{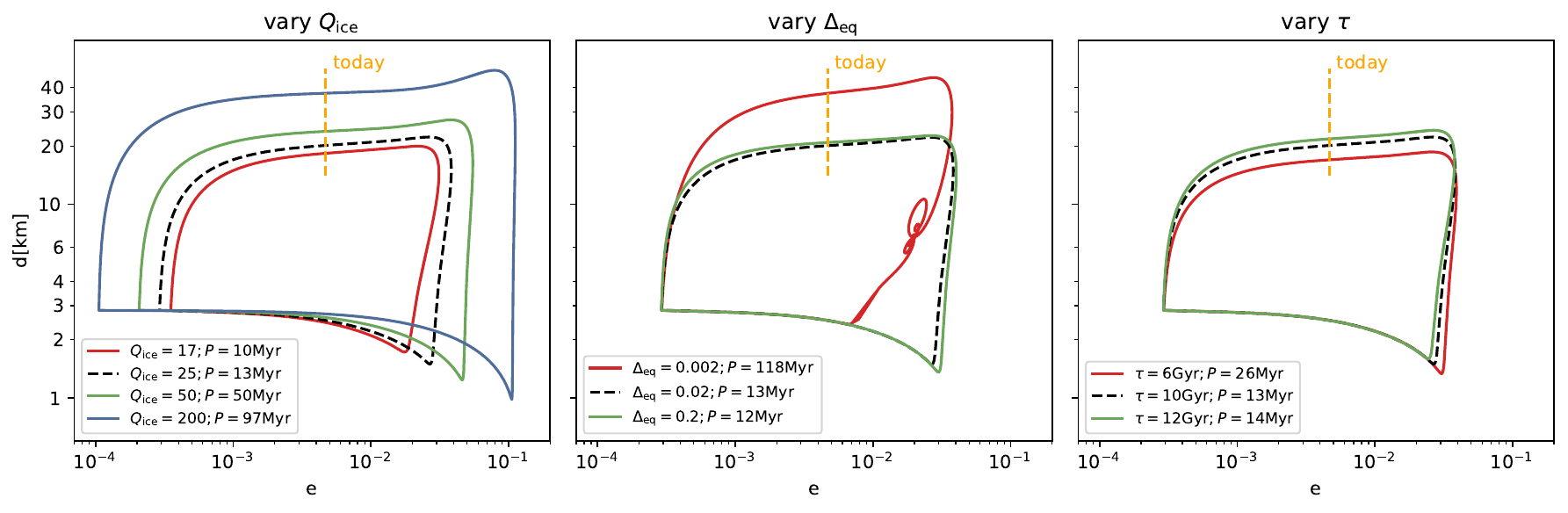}
\caption{{\bf Limit Cycle for Different Parameters}. 
 In each of the three panels, 
we vary one parameter, while keeping the other parameters at their fiducial values. 
The insets show the values of the parameter that
is varied, as well as the period of the limit
cycle for that parameter value.
}
\label{fig:varyAll}
%\vspace{1cm} 
\end{figure*}
%%%%%%%%%%%%%%%%%%%%%%%%%%%%%%%%%%

In the resonant libration stage,   $d$ slowly grows towards its resonant value  $d_{\rm res}=2.8$km, while the
heating remains high: $H\sim 230$GW throughout most of this stage. 
This behavior may be understood by examining the $H$ and $C$ curves
in the left panel of Figure \ref{fig:heating}. 
As the system follows the $C$ curve towards $d_{\rm res}$, which has magnitude
$C(d=d_{\rm res})=230$GW, heating very nearly balances cooling. 
If $e$ were to be held artificially fixed, then the system would remain 
in thermal equilibrium, with $H=C$, and a little to the left of the resonant peak at $d_{\rm res}$. 
This is a stable equilibrium: an increase in $d$ enhances heating relative to cooling, which
melts the shell, decreasing $d$. In contrast, the equilibrium just to the right of the resonant peak is unstable.
Now, with $e$ allowed to evolve, the fact that the heating rate of $\sim 230$GW exceeds $H_{\rm eq}$
forces the moons to diverge, lowering the resonantly forced $e$, which lowers the red curve in Figure \ref{fig:heating}.  Eventually, the red curve can no longer reach the blue curve, which occurs
when $e=e_{\rm min}=0.0003$. 
When that happens, the resonant libration stage ends. Heating cannot compete with cooling, 
causing
 $d$ to increase beyond $d_{\rm res}$. That makes heating extremely small because it
can no longer take advantage of the resonant peak. The system thus re-enters the freezing stage.
At the beginning of the freezing stage, with $H\approx 0$, 
the cooling rate shrinks from 230GW in inverse proportion to the growing $d$ (Fig. \ref{fig:heat}), 
crossing
 through 32GW
today.  The fact that the  heating is very small forces the moons to converge, and the cycle repeats.

\section{Discussion}

We  presented a simple model for the  thermal and tidal dynamics of Enceladus.
With our chosen values 
for the three uncertain parameters in the model 
($\Q$,  $\Delta_{\rm eq}$, and $\tau$), we found 
that Enceladus does not remain at its  equilibrium point, but instead follows a limit
cycle in the $e$-$d$ plane. 
We consider most uncertain input physics in the model to be the resonant locking hypothesis, as modelled
by equation (\ref{eq:trl})
	 for Saturn's torque on Enceladus. 

Many of Enceladus's observed properties
are consistent with it currently being in the freezing phase of its limit cycle, including
$e_{\rm obs}$, $\gamma_{\rm obs}$, $d_{\rm obs}$, and $C_{\rm obs}$.
As hinted at the end of the previous section, 
the reason Enceladus escapes from  equilibrium  
is that the equilibrium is thermally unstable
\citep{2022Icar..37314769S}. In the left panel of Figure \ref{fig:heating}, global equilibrium occurs
where the blue and green curves intersect.  At that position, the red curves fall more steeply than 
the blue, implying that an increase in $d$ favors cooling over heating, which increases $d$ even more. 
Previous studies have considered limit cycles for Io 	 \citep{1986Icar...66..341O} and Enceladus
\citep{2008Icar..198..178M,2014Icar..235...75S}, 
but driven by a different mechanism, in which the moon's dissipative properties
were modelled in a more ad hoc way.

A difficulty with our   model is that
Enceladus's current migration rate appears to be too fast.  
 In the right panel of Figure \ref{fig:delta}, 
the slope of the orange curve   at the time marked ``today'' yields that
 $\dot{n}/n \approx -10/\tau=-1/{\rm Gyr}$, which is around ten times
 faster than that of \cite{2020NatAs...4.1053L}. Whether this is truly a
 discrepancy is not certain, as Lainey et al's value relies on modelling of the tides
 on both Saturn and on Enceladus,
 in a different way than we do, and is complicated by the interaction between
 Enceladus and Dione. But if there is a discrepancy, our model might
 be adjusted in a number of ways to resolve it:  for example,
 by including a more elaborate model for Saturn's tidal torque, or by including
 tidal damping of Dione.  In Figure \ref{fig:delta}, one sees that  Enceladus's migration rate slows 
 very
 soon after ``today.'' A horizontal line in that plot corresponds
 to a migration time of $\tau =10$Gyr. Hence the migration rate
 becomes
  close to Lainey et al's value  in the very near future.  It  is therefore possible that a modest adjustment of our model 
 would suffice to resolve the difficulty.

We find that the behavior of the limit cycle is not greatly sensitive to the
values of the parameters.  
In Figure \ref{fig:varyAll} we show how the limit
cycle changes when $\Q$, $\Delta_{\rm eq}$, and
$\tau$ are varied.
From the left panel we see that as $\Q$
is varied from 17 to 200, the
 present-day thickness of the ice shell increases from $18$km to 37km;  the 
minimum shell thickness decreases from 1.7km to 1km; and
the
period of a  cycle increases from 10 to
  97 Myr. We also find that for $\Q\lesssim 16$, the limit cycle goes
  away, and Enceladus reaches
  its global equilibrium.
From the middle panel, we see that one may vary 
$\Delta_{\rm eq}$ by two orders of magnitude, without
much effect on the cycle.  In the right panel, we
vary $\tau$ by a factor of 2, and also find not
much of an effect on the cycle. However, 
if either $\tau\lesssim 5$Gyr, or $\tau\gtrsim 13$Gyr, 
Enceladus does not evolve away from its
initial state, i.e., the global equilibrium state is stable.
The reason for this sensitivity may be understood 
from the left panel of Figure \ref{fig:heating}:  changing $\tau$ moves the
$H_{\rm eq}$  line up and down. 
If it is moved  too far up, the global equilibrium 
is to the left of the resonant peak, and hence is stable. 
And if it is moved too far down, the global equilibrium
occurs where the red and blue curves are nearly parallel
to each other, indicating neutral stability.
Nonetheless, by adjusting the initial value of 
$d$, we have found that limit cycles persist
for $\tau$ up to 18Gyr.
We leave a more thorough investigation of parameter space to the future.

In conclusion, we have shown that Enceladus is likely
not in equilibrium.
Instead, its semimajor axis, eccentricity, and shell thickness, are tracing out a limit cycle. 
In addition to the evidence presented above, 
three further lines of evidence support
this conclusion:
\begin{enumerate}
\item 
The expression for the heating rate (eq. \ref{eq:hed}) shows that
$H\approx (25/ \Q)$GW today, after inserting
the observed  $d_{\rm obs}$ and $e_{\rm obs}$.
Thus one would require an unreasonably low $\Q$  for
the heating to be comparable to
$C_{\rm obs}\sim 20$-$30$GW. 
 \cite{2022Icar..37314769S} have argued similarly.
This strongly suggests that
heating is less that $C_{\rm obs}$, implying that
Enceladus is currently in the freezing stage, in which its ice shell is thickening. 

\item  
The cracks  seen in the south pole region are a natural consequence  of a thickening
ice shell  \citep[e.g.,][]{2022GeoRL..4994421R}.  
As the shell thickens, the increased volume occupied by ice relative to water overpressurizes
the ocean, and the resulting stresses are sufficiently strong to
crack Enceladus's ice shell.
By contrast, tidal stresses are likely too weak to crack the shell, 
as they are
an order of magnitude weaker.

\item
The shapes of craters on Enceladus are more relaxed than they should be based on present conditions.  
But the shapes have been explained by a past episode of extreme heating, in which the heat flux exceeded 150 mW m$^{-2}$, or 120GW averaged over the moon \citep{2012GeoRL..3917204B}. 
Such a high heat flux occurs in our model throughout the resonant libration stage, when $C\approx 230$GW. 

\end{enumerate}

One might wonder whether Europa exhibits behavior similar to Enceladus.  An important difference between the two moons is that Europa's ice shell cannot experience resonant libration, whatever
its thickness
%.  
%That is because Europa's shell is in the ``soft'' limit, which implies that 
%its libration frequency does not vary with $d$
 \citep[][Lithwick 2025]{2010Icar..209..631G}.
 As a result, Europa cannot experience a limit
 cycle of the type described in this paper. 

\begin{acknowledgments}
We thank Carolyn Porco for helpful discussions.
Y.L. acknowledges NASA grant 80NSSC23K1262.
\end{acknowledgments}

\clearpage

\appendix
\section{Derivation of Simplified Orbital Equations}
\label{sec:appendix}

We derive equations (\ref{eq:torque_2})--(\ref{eq:power_2}), starting from equations (\ref{eq:torqueeq})--(\ref{eq:eorb}), 
together with our model for the resonance locking torque (eqs. \ref{eq:omdot}--\ref{eq:trl}). 
We first
define
the nominal angular momenta of the two moons as
\be
L_{\rm nom}&=&m(GM_S)^{2/3}\omega^{-1/3} \\
L_{2,\rm nom}&=&m_2(GM_S)^{2/3}(\omega/2)^{-1/3}
\ee
We  expand the angular momenta and energies
 to first order in $\Delta$, $\Delta_2$ (as defined in eq. \ref{eq:delta1} \& \ref{eq:delta2})  and in $e^2$:
\be
L
&\approx& L_{\rm nom}\left( 1+{1\over 3}{\Delta}-{e^2\over 2} \right)
\label{eq:lnom}
\\
L_2
&\approx& L_{2,\rm nom}\left(1+{1\over 3}\Delta_2  \right) 
\\
E
&\approx& \omega L_{\rm nom}\left( -{1\over 2}+{1\over 3}{\Delta} \right) \\
E_2
&\approx&  {\omega\over 2}L_{2,\rm nom}\left( -{1\over 2}+{1\over 3}\Delta_2   \right)
\label{eq:e2nom}
\ee
We then write the torque equation  (eq. \ref{eq:torqueeq}) as follows:
\be
T-\dot{L}_{\rm nom}-\dot{L}_{2,\rm nom}&=&{d\over dt}\left(L-L_{\rm nom}+L_2-L_{2,\rm nom}   \right) 
\ee
which from equations (\ref{eq:lnom})--(\ref{eq:e2nom})
and equation (\ref{eq:omdot}) approximates to 
\be
T-{1\over 3\tau}\left(L_{\rm nom}+L_{2,\rm nom}\right)\approx \left(L_{\rm nom}+L_{2,\rm nom}\right){1\over 1+\epsilon}\left(
\epsilon({1\over 3}\dot{\Delta}-e\dot{e})+{1\over 3}\dot{\Delta}_2
\right) \label{eq:torque2}
\ee
after defining
\be
\epsilon&\equiv& {L_{\rm nom}\over L_{2,\rm nom}} 
= {m\over  2^{1/3} m_2} 
\ee
We have  
assumed that the timescale of variation of 
our new variables ($\Delta$, $\Delta_2$, and $e$) is
much shorter than $\tau$ in equation (\ref{eq:omdot}), 
which enables us to drop, e.g., 
$\dot{L}_{\rm nom}\Delta$ in comparison with
$L_{\rm nom}\dot{\Delta}$. 

For the power equation (eq. \ref{eq:power}), we proceed  similarly.   But we first subtract $\omega\times $  (eq. \ref{eq:torqueeq}), which gives
\be
H
 &=& \omega\dot{L}+\omega\dot{L}_2-\dot{E}-\dot{E}_2
\ee
after dropping the small term $(n-\omega)T$.
Inserting the approximate forms for $L$ and $E$, and moving the time-derivative of the nominal pieces
of those approximate forms to the left-hand-side of the equation, yields
\be
H
-{\omega\over 3\tau}\left(L_{\rm nom}+L_{2,\rm nom}\right)+{\omega\over 3\tau} L_{\rm nom}+{\omega\over 2\cdot 3\tau}L_{2,\rm nom} = -\omega L_{\rm nom}e\dot{e}  + \omega L_{2,\rm nom}{1\over 6}\dot{\Delta}_2
\ee
and then after multiplying  through:
\be
{H(1+\epsilon)\over \omega \left(L_{\rm nom}+L_{2,\rm nom}\right)}-{1\over 6\tau}=  -\epsilon e\dot{e}  +{1\over 6}\dot{\Delta}_2 \label{eq:power2b}
\ee
 Equations (\ref{eq:torque2}) and (\ref{eq:power2b}) are the  equations that we integrate numerically
 in the body of the paper. The terms in these equations have transparent physical meanings. For example,
 the term on the right-hand-side of equation (\ref{eq:torque2}) that is  proportional to $\epsilon$ is the rate of change of Enceladus's orbital angular momentum minus what it would have been on its nominal orbital expansion. And the other term is the same for Dione. 
 
 Finally, it is convenient to rewrite those two equations by introducing two new constants:
 \be
 \Delta_{\rm eq}&\equiv& \left( {3\tau T\Delta^2\over L_{\rm nom}+L_{2,\rm nom}} \right)^{1/2}  \label{eq:deleqdef} \\
H_{\rm eq}&\equiv& {\omega \over 6\tau(1+\epsilon)} \left(L_{\rm nom}+L_{2,\rm nom}\right)  \label{eq:ldef}
\ee
where  the first expression is constant
because $T\Delta^2=$ is constant (eq. \ref{eq:tdef}).
 Equations (\ref{eq:torque2}) and (\ref{eq:power2b})  then turn into the equations given in the body of the paper
  (eqs. \ref{eq:torque_2}--\ref{eq:power_2}).

\bibliography{encel}{}

\begin{thebibliography}{}
\expandafter\ifx\csname natexlab\endcsname\relax\def\natexlab#1{#1}\fi
\providecommand{\url}[1]{\href{#1}{#1}}
\providecommand{\dodoi}[1]{doi:~\href{http://doi.org/#1}{\nolinkurl{#1}}}
\providecommand{\doeprint}[1]{\href{http://ascl.net/#1}{\nolinkurl{http://ascl.net/#1}}}
\providecommand{\doarXiv}[1]{\href{https://arxiv.org/abs/#1}{\nolinkurl{https://arxiv.org/abs/#1}}}

\bibitem[{{Bland} {et~al.}(2012){Bland}, {Singer}, {McKinnon}, \& {Schenk}}]{2012GeoRL..3917204B}
{Bland}, M.~T., {Singer}, K.~N., {McKinnon}, W.~B., \& {Schenk}, P.~M. 2012, \grl, 39, L17204, \dodoi{10.1029/2012GL052736}

\bibitem[{{{{C}}adek} {et~al.}(2016){{{C}}adek}, {Tobie}, {Van{\^A} Hoolst}, {Mass{\'e}}, {Choblet}, {Lef{\`e}vre}, {Mitri}, {Baland}, {B{\v{e}}hounkov{\'a}}, {Bourgeois}, \& {Trinh}}]{2016GeoRL..43.5653C}
{{{C}}adek}, O., {Tobie}, G., {Van{\^A} Hoolst}, T., {et~al.} 2016, \grl, 43, 5653, \dodoi{10.1002/2016GL068634}

\bibitem[{{{\'C}uk} {et~al.}(2024){{\'C}uk}, {El Moutamid}, {Lari}, {Neveu}, {Nimmo}, {Noyelles}, {Rhoden}, \& {Saillenfest}}]{2024SSRv..220...20C}
{{\'C}uk}, M., {El Moutamid}, M., {Lari}, G., {et~al.} 2024, \ssr, 220, 20, \dodoi{10.1007/s11214-024-01049-2}

\bibitem[{{Fuller} {et~al.}(2016){Fuller}, {Luan}, \& {Quataert}}]{2016MNRAS.458.3867F}
{Fuller}, J., {Luan}, J., \& {Quataert}, E. 2016, \mnras, 458, 3867, \dodoi{10.1093/mnras/stw609}

\bibitem[{{Goldreich} \& {Mitchell}(2010)}]{2010Icar..209..631G}
{Goldreich}, P.~M., \& {Mitchell}, J.~L. 2010, \icarus, 209, 631, \dodoi{10.1016/j.icarus.2010.04.013}

\bibitem[{{Hsu} {et~al.}(2015){Hsu}, {Postberg}, {Sekine}, {Shibuya}, {Kempf}, {Hor{\'a}nyi}, {Juh{\'a}sz}, {Altobelli}, {Suzuki}, {Masaki}, {Kuwatani}, {Tachibana}, {Sirono}, {Moragas-Klostermeyer}, \& {Srama}}]{2015Natur.519..207H}
{Hsu}, H.-W., {Postberg}, F., {Sekine}, Y., {et~al.} 2015, \nat, 519, 207, \dodoi{10.1038/nature14262}

\bibitem[{{Iess} {et~al.}(2014){Iess}, {Stevenson}, {Parisi}, {Hemingway}, {Jacobson}, {Lunine}, {Nimmo}, {Armstrong}, {Asmar}, {Ducci}, \& {Tortora}}]{2014Sci...344...78I}
{Iess}, L., {Stevenson}, D.~J., {Parisi}, M., {et~al.} 2014, Science, 344, 78, \dodoi{10.1126/science.1250551}

\bibitem[{{Lainey} {et~al.}(2012){Lainey}, {Karatekin}, {Desmars}, {Charnoz}, {Arlot}, {Emelyanov}, {Le Poncin-Lafitte}, {Mathis}, {Remus}, {Tobie}, \& {Zahn}}]{2012ApJ...752...14L}
{Lainey}, V., {Karatekin}, {\"O}., {Desmars}, J., {et~al.} 2012, \apj, 752, 14, \dodoi{10.1088/0004-637X/752/1/14}

\bibitem[{{Lainey} {et~al.}(2020){Lainey}, {Casajus}, {Fuller}, {Zannoni}, {Tortora}, {Cooper}, {Murray}, {Modenini}, {Park}, {Robert}, \& {Zhang}}]{2020NatAs...4.1053L}
{Lainey}, V., {Casajus}, L.~G., {Fuller}, J., {et~al.} 2020, Nature Astronomy, 4, 1053, \dodoi{10.1038/s41550-020-1120-5}

\bibitem[{{Le Gall} {et~al.}(2017){Le Gall}, {Leyrat}, {Janssen}, {Choblet}, {Tobie}, {Bourgeois}, {Lucas}, {Sotin}, {Howett}, {Kirk}, {Lorenz}, {West}, {Stolzenbach}, {Mass{\'e}}, {Hayes}, {Bonnefoy}, {Veyssi{\`e}re}, \& {Paganelli}}]{2017NatAs...1E..63L}
{Le Gall}, A., {Leyrat}, C., {Janssen}, M.~A., {et~al.} 2017, Nature Astronomy, 1, 0063, \dodoi{10.1038/s41550-017-0063}

\bibitem[{{Meyer} \& {Wisdom}(2007)}]{2007Icar..188..535M}
{Meyer}, J., \& {Wisdom}, J. 2007, \icarus, 188, 535, \dodoi{10.1016/j.icarus.2007.03.001}

\bibitem[{{Meyer} \& {Wisdom}(2008)}]{2008Icar..198..178M}
---. 2008, \icarus, 198, 178, \dodoi{10.1016/j.icarus.2008.06.012}

\bibitem[{{Murray} \& {Dermott}(1999)}]{1999ssd..book.....M}
{Murray}, C.~D., \& {Dermott}, S.~F. 1999, {Solar System Dynamics}, \dodoi{10.1017/CBO9781139174817}

\bibitem[{{Nimmo} {et~al.}(2023){Nimmo}, {Neveu}, \& {Howett}}]{2023SSRv..219...57N}
{Nimmo}, F., {Neveu}, M., \& {Howett}, C. 2023, \ssr, 219, 57, \dodoi{10.1007/s11214-023-01007-4}

\bibitem[{{Ojakangas} \& {Stevenson}(1986)}]{1986Icar...66..341O}
{Ojakangas}, G.~W., \& {Stevenson}, D.~J. 1986, \icarus, 66, 341, \dodoi{10.1016/0019-1035(86)90163-6}

\bibitem[{{Park} {et~al.}(2024){Park}, {Mastrodemos}, {Jacobson}, {Berne}, {Vaughan}, {Hemingway}, {Leonard}, {Castillo-Rogez}, {Cockell}, {Keane}, {Konopliv}, {Nimmo}, {Riedel}, {Simons}, \& {Vance}}]{2024JGRE..12908054P}
{Park}, R.~S., {Mastrodemos}, N., {Jacobson}, R.~A., {et~al.} 2024, Journal of Geophysical Research (Planets), 129, e2023JE008054, \dodoi{10.1029/2023JE008054}

\bibitem[{{Peale} \& {Cassen}(1978)}]{1978Icar...36..245P}
{Peale}, S.~J., \& {Cassen}, P. 1978, \icarus, 36, 245, \dodoi{10.1016/0019-1035(78)90109-4}

\bibitem[{{Porco} {et~al.}(2014){Porco}, {DiNino}, \& {Nimmo}}]{2014AJ....148...45P}
{Porco}, C., {DiNino}, D., \& {Nimmo}, F. 2014, \aj, 148, 45, \dodoi{10.1088/0004-6256/148/3/45}

\bibitem[{{Porco} {et~al.}(2006){Porco}, {Helfenstein}, {Thomas}, {Ingersoll}, {Wisdom}, {West}, {Neukum}, {Denk}, {Wagner}, {Roatsch}, {Kieffer}, {Turtle}, {McEwen}, {Johnson}, {Rathbun}, {Veverka}, {Wilson}, {Perry}, {Spitale}, {Brahic}, {Burns}, {Del Genio}, {Dones}, {Murray}, \& {Squyres}}]{2006Sci...311.1393P}
{Porco}, C.~C., {Helfenstein}, P., {Thomas}, P.~C., {et~al.} 2006, Science, 311, 1393, \dodoi{10.1126/science.1123013}

\bibitem[{{Rudolph} {et~al.}(2022){Rudolph}, {Manga}, {Walker}, \& {Rhoden}}]{2022GeoRL..4994421R}
{Rudolph}, M.~L., {Manga}, M., {Walker}, M., \& {Rhoden}, A.~R. 2022, \grl, 49, e2021GL094421, \dodoi{10.1029/2021GL094421}

\bibitem[{{Shao} \& {Nimmo}(2022)}]{2022Icar..37314769S}
{Shao}, W.~D., \& {Nimmo}, F. 2022, \icarus, 373, 114769, \dodoi{10.1016/j.icarus.2021.114769}

\bibitem[{{Shoji} {et~al.}(2014){Shoji}, {Hussmann}, {Sohl}, \& {Kurita}}]{2014Icar..235...75S}
{Shoji}, D., {Hussmann}, H., {Sohl}, F., \& {Kurita}, K. 2014, \icarus, 235, 75, \dodoi{10.1016/j.icarus.2014.03.006}

\bibitem[{{Thomas} {et~al.}(2016){Thomas}, {Tajeddine}, {Tiscareno}, {Burns}, {Joseph}, {Loredo}, {Helfenstein}, \& {Porco}}]{2016Icar..264...37T}
{Thomas}, P.~C., {Tajeddine}, R., {Tiscareno}, M.~S., {et~al.} 2016, \icarus, 264, 37, \dodoi{10.1016/j.icarus.2015.08.037}

\bibitem[{{Van Hoolst} {et~al.}(2013){Van Hoolst}, {Baland}, \& {Trinh}}]{2013Icar..226..299V}
{Van Hoolst}, T., {Baland}, R.-M., \& {Trinh}, A. 2013, \icarus, 226, 299, \dodoi{10.1016/j.icarus.2013.05.036}

\end{thebibliography}
\bibliographystyle{aasjournal}

\end{document}